\documentclass[twocolumn,showpacs,preprintnumbers,amsmath,amssymb]{revtex4}

\usepackage{graphicx}
\usepackage{dcolumn}
\usepackage{bm}

\begin{document}


\title{Boundary effects on energy dissipation in a cellular automaton model}

\author{Wei Zhang}
 \email{tzwphys@jnu.edu.cn; wzhang2007065@gmail.com}
\author{Wei Zhang}%
 \email{twzhang@jnu.edu.cn}
\affiliation{%
Department of Physics, Jinan University, Guangzhou 510632, China\\
}%

\date{\today}

\begin{abstract}
In this paper, we numerically study energy dissipation caused by
traffic in the Nagel-Schreckenberg (NaSch) model with open boundary
conditions (OBC). Numerical results show that there is a
nonvanishing energy dissipation rate $E_d$, and no true free-flow
phase exists in the deterministic and nondeterministic NaSch models
with OBC. In the deterministic case, there is a critical value of
the extinction rate $\beta _{cd}$ below which $E_d$ increases with
increasing $\beta $, but above which $E_d$ abruptly decreases in the
case of the speed limit $v_{\max }\geqslant 3.$ However, when
$v_{\max }\leqslant \ 2,$ no discontiguous change in $E_d$ occurs.
In the nondeterministic case, the dissipated energy has two
different contributions: one coming from the randomization, and one
from the interactions, which is the only reason for dissipating
energy in the deterministic case. The relative contributions of the
two dissipation mechanisms are presented in the stochastic NaSch
model with OBC. Energy dissipation rate $E_d$ is directly related to
traffic phase. Theoretical analyses give an agreement with numerical
results in three phases (low-density, high-density and maximum
current phase) for the case $v_{\max }=1.$
\end{abstract}

\maketitle

\section{\label{sec:level1}INTRODUCTION}

In the last decades, traffic problems have attracted much attention
of a community of physicists because of the observed nonequilibrium
phase transitions and various nonlinear dynamical phenomena. A
number of traffic models have been proposed to investigate the
dynamical behavior of the traffic flow, including fluid dynamical
models, gas-kinetic models, car-following models and cellular
automata (CA) models\cite{1,2,3,4}. These dynamical approaches
represented complex physical phenomena of traffic flow among which
are hysteresis, synchronization, wide moving jams, and phase
transitions, etc. Among these models, the cellular automata
approaches can be used very efficiently for computers to perform
simulation\cite {1,4,5,6,7,8,9,10,11,12,13,14}. The
Nagel-Schreckenberg (NaSch) model is a basic CA models describing
one-lane traffic flow\cite{5}. Based on the NaSch model, many CA
models have succeeded in modeling a wide variety of properties of
vehicular traffic\cite{1,4,7,8,9,10,11,12,13,14}.

On the other hand, the problems of traffic jams, environmental
pollution and energy dissipation caused by traffic have become more
and more significant in modern society. Financial damage from
traffic due to energy dissipation and environmental pollution is
huge every year. In accordance with previously reported results in
Refs.[15], more than 20\% fuel consumption and air pollution is
caused by impeded and ''go and stop'' traffic. Most recently, the
problem of energy dissipation in traffic system has been
investigated in the framework of car following model, city traffic
model and NaSch model with periodic boundary conditions (PBC),
respectively\cite {16,17,18,19,20}. And analytical expressions for
energy dissipation have also been provided in the nondeterministic
NaSch model with PBC in the case of $v_{\max }=1$ and free-flow
state\cite{20}. However, the effects of boundary condition on energy
dissipation have not been discussed yet.

The most significant difference between systems with open and
periodic boundary conditions is the vehicle density $\rho $. In a
periodic system, which has no maximum current phase, vehicle density
is considered as an adjustable parameter. In systems with open
boundary conditions (OBC),
however, there are two adjustable parameters, namely the injection rate $%
\alpha $ and the extinction rate $\beta $, and the vehicle density
$\rho $ is only a derived parameter. Compared with periodic systems,
it implies that open systems show a different behaviour of
quantities such as the global density, the current, the density
profile and even the microscopic structure of traffic
phase\cite{21,22,23,24,25,26,27,28,29,30}. Therefore, energy
dissipation in the CA model with OBC, which is relevant to many
realistic situation in traffic, should be further investigated.

In this paper, we investigate the energy dissipation rate within the
framework of the deterministic and nondeterministic NaSch model with
OBC. The behaviours of the energy dissipation rate in different
traffic phase are distinct. Theoretical analyses are presented in
low-density, high-density and maximum current phase in the case of
$v_{\max }=1$. The influences of the speed limit $v_{\max }$ on the
energy dissipation are also investigated. Energy dissipation caused
by braking is related not only to the velocity of vehicles, but also
to the headway distribution. Thus, the behaviour of the energy
dissipation rate is more complex than simpler quantities, and should
be further investigated. The results of this article may lead to a
profound understanding of some features of traffic system or may
provide schemes for reducing energy dissipation of the existing
traffic network.

The paper is organized as follows. Section II is devoted to the
description of the model and the definition of energy dissipation
rate. In section III, the numerical studies are given, and the
influences of the injection and extinction rate on energy
dissipation rate are considered. And theoretical analyses are
presented in the special case $v_{\max }=1$. Finally, the
conclusions are given in section IV.

\section{MODEL AND ENERGY DISSIPATION}

Our investigations are based on a one dimensional cellular automaton
model introduced by Nagel and Schreckenberg. The model is defined on
a single lane road consisting of $L$ cells of equal size numbered by
$i=1,$ $2,$ $\cdots ,$ $L$ and the time is discrete. Each site can
be either empty or occupied by a car with the speed $v=0,$ $1,$
$2,\cdots $ $,$ $v_{\max }$, where $v_{\max }$ is the speed limit.
Let $x(i,t)$ and $v(i,t)$ denote the position and the velocity of
the $i$th car at time $t$, respectively. The number of empty
cells in front of the $i$th vehicle is denoted by $d(i,t)=x(i+1,t)-x(i,t)-1$%
. The following four steps for all cars update in parallel with
periodic boundary.

(1) Acceleration:

$v(i,t+1/3)\rightarrow \min [v(i,t)+1,v_{\max }];$

(2) Slowing down:

$v(i,t+2/3)\rightarrow \min [v(i,t+1/3),d(i,t)];$

(3) Stochastic braking:

$v(i,t+1)\rightarrow \max [v(i,t+2/3)-1,0]$ with the probability
$p;$

(4) Movement: $x(i,t+1)\rightarrow x(i,t)+v(i,t+1).$

Open systems are characterized by the injection rate $\alpha $ and
the
extinction rate $\beta $, which means by the probability $\alpha $ and $%
\beta $ that a vehicle moves into and out of the system. In this
paper, open boundary conditions are defined according
to\cite{28,29}. At site $i=0$ which means out of the system, a
vehicle with speed $v=v_{\max }$ is created with probability $\alpha
$. The vehicle immediately moves forward in accordance with the
NaSch rule. If the site $i=1$ is occupied by a car, the injected
vehicle at site $i=0$ is deleted. At $i=L+1$ a ''block'' occurs with
probability $1-\beta $ and causes a slowing down of the vehicles at
the end of the system. Otherwise, a vehicle may leave freely from
the end of the system.

It should be mentioned that for $v_{\max }=1$ the model above in the
nondeterministic case is different from parallel updated asymmetric
exclusion process (ASEP) with open boundary
conditions\cite{21,22,23}. In the ASEP model, if the site $L$ is
occupied then the particle on that site exits with probability
$\beta $, irrespective of their velocity. In the model above,
however, the extinction of the vehicle at site $L$ depends not only
on the probability $\beta $, but also on the braking probability
$p.$ Even if the vehicle with $v=1$ is at site $L$ and there is no
''block'' at site $L+1$, it may fail to exit because of being
randomly delayed. This difference may influence the current and
phase transition, which will be analysed in the following section.
In the deterministic case, however, the model above with $v_{\max
}=1$ is identical with parallel updated ASEP with OBC, for there is
no stochastic delay.

The kinetic energy of the vehicle with the velocity $v$ is $mv^2/2$, where $%
m $ is the mass of the vehicle. When braking the energy is lost. Let
$E_d$ denotes energy dissipation rate per time step per vehicle. For
simple, we neglect rolling and air drag dissipation and other
dissipation such as the energy needed to keep the motor running
while the vehicle is standing in our analysis, i.e., we only
consider the energy lost caused by speed-down. The dissipated energy
of $i$th vehicle from time $t-1$ to $t$ is defined by\cite {20}

\[
e(i,t)=%
{\frac m2\left[ v^2(i,t-1)-v^2(i,t)\right] \quad \text{for }v(i,t)<v(i,t-1) \atopwithdelims\{. 0\qquad \qquad \qquad \qquad \qquad ~~\text{for }v(i,t)\geqslant v(i,t-1).}%
\qquad \left( 1\right)
\]
Thus, the energy dissipation rate

\[
E_d=\frac 1T\frac 1N\sum_{t=t_0+1}^{t_0+T}\sum_{i=1}^Ne(i,t),\qquad
\left( 2\right)
\]
where $N$ is the number of vehicles in the system and $t_0$ is the
relaxation time, taken as $t_0=10^5$. In this model, the particles
are ''self-driven'' and the kinetic energy increases in the
acceleration step. In the stationary state, the value of the
increased energy while accelerating is equivalent to that of the
dissipated energy caused by speed-down, and the kinetic energy is
constant in the system. In the simulation, the system size $L=1000$
is selected, and the results are obtained by averaging over 20
initial configurations and $10^4$ time steps after discarding $10^5$
initial transient states.

\section{NUMERICAL RESULTS}

\subsection{Effects of the boundary conditions on energy dissipation in the
deterministic case}

First, we investigate the influences of the boundary conditions on
energy
dissipation in the deterministic NaSch model with the maximum velocity $%
v_{\max }=5$. In the deterministic case, the stochastic braking is
not considered, i.e., $p=0$. Figure 1 shows the energy dissipation
rate $E_d$ as a function of the extinction rate $\beta $ with
different values of the injection rate $\alpha $. As shown in Fig.
1, there is a critical value of the extinction rate $\beta _{cd}$
below which $E_d$ increases with the increase of the rate $\beta $,
but above which $E_d$ abruptly decreases. The position of $\beta
_{cd}$ shifts towards a high value of $\beta $ with increasing the
injection rate $\alpha $. The astonishing result is that there is a
nonvanishing energy dissipation, even though in low-density phase,
which means that there is no ''true'' free-flow phase.

\begin{figure}
\includegraphics[height=5cm]{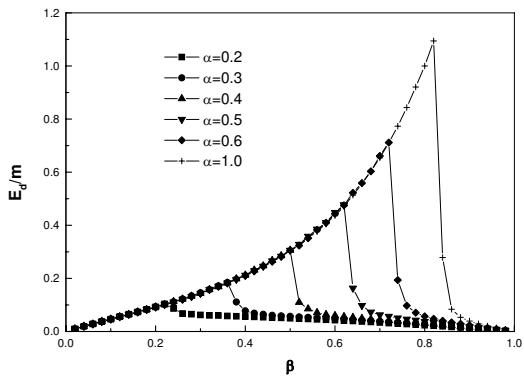}
\caption{\label{fig:epsart} Energy dissipation rate $E_d$ (scaled by
$m$ ) as a function of the extinction rate $\beta $ in the
deterministic NaSch model with $v_{\max }=5$ for various values of
the injection rate $\alpha $.}
\end{figure}

When $\beta <\beta _{cd},$ the state of the system is high-density
phase. As the extinction rate $\beta $ increases, the kinetic energy
possessed by vehicles increases because of the increase of the mean
vehicle velocity, and the dissipated energy while braking increases.
When $\beta >\beta _{cd},$ the state of the system is low-density
phase in which the distance-headways is larger and the interaction
between vehicles is weaker than that in the high-density phase. The
interaction is the only reason for dissipating energy in the
deterministic case. Consequently, $E_d$ decreases abruptly when the
transition from high-density to low-density phase occurs. Because of
boundary effects there are vehicular interactions even at
low-density phase. Therefor, there is a nonvanishing energy
dissipation rate $E_d$, and no ''true'' free-flow phase exists in
the deterministic NaSch model with open boundary conditions.

Energy dissipation rate $E_d$ in the case of $v_{\max }=\ 3$ and 4
show similar behaviour to that for $v_{\max }=5,$ and the position
of $\beta _{cd} $ shifts towards a low value of $\beta $ with
increasing the speed
limit $v_{\max },$ as shown in Fig. 2. But in the case of $v_{\max }=2$ and $%
1$, there is no critical value of the extinction rate $\beta _{cd}$,
i.e., no discontinuous change in $E_d$ occurs. According to
previously reported results in Refs.[19], in the case of $v_{\max
}<3$, the state of a system
with injection rate $\alpha =1$ is the jamming phase, while in the case of $%
v_{\max }\geqslant \ 3$, the jamming state exists in the region of
low value of $\beta $, and low-density lies in the region of high
value of $\beta .$ Thus, there is no discontinuous change in $E_d$
in the case of $v_{\max }=2$ and $1$ for no phase transition occurs.
Near $\beta =0$, energy dissipation rate $E_d$ is independent of the
speed limit $v_{\max }$, and $E_d$ increases linearly with the
increase of $\beta $. In the case of $\alpha =1$ and $\beta
\rightarrow 0$, the maximum velocity which vehicles can move is 1,
so the speed limit has no influences on energy dissipation.

\begin{figure}
\includegraphics[height=5cm]{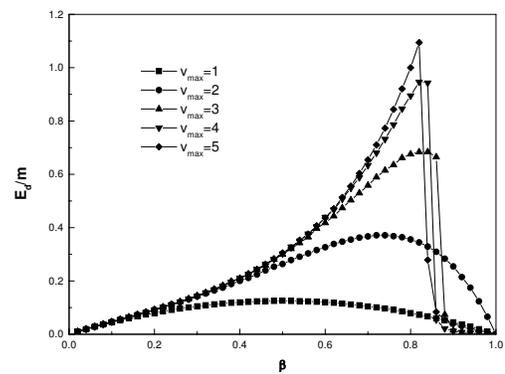}
\caption{\label{fig:epsart} Energy dissipation rate $E_d$ (scaled by
$m$ ) as a function of the extinction rate $\beta $ in the
deterministic NaSch model in the case of $\alpha =1.0$ for various
values of the speed limit $v_{\max }$.}
\end{figure}

\begin{figure}
\includegraphics[height=5cm]{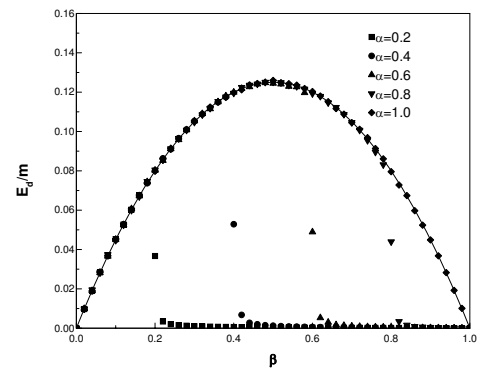}
\caption{\label{fig:epsart} Energy dissipation rate $E_d$ (scaled by
$m$ ) as a function of the extinction rate $\beta $ in the
deterministic NaSch model with $v_{\max }=1$ for various values of
the injection rate $\alpha $. Symbol data are obtained from computer
simulations, and solid line corresponds to analytic results of the
formula (6).}
\end{figure}

In the special case of $v_{\max }=1,$ energy dissipation rate $E_d$
is proportional to the mean density of ''go and stop'' vehicles per
time step. The mean density of ''go and stop'' vehicles is defined
in the following way:

\[
\rho _{gs}=\frac 1N\frac
1T\sum_{t=t_0+1}^{t_0+T}\sum_{i=1}^Nn(i,t)\left[ 1-n(i,t+1)\right]
,\qquad \qquad (3)
\]
where $n(i,t)=0$ for stopped cars and $n(i,t)=1$ for moving cars at time $t$%
, and $t_0$ is the relaxation time as mentioned in the section II.
And energy dissipation rate $E_d$ in the case of $v_{\max }=1$ can
be written as
\[
E_d=\frac 12m\rho _{gs}.\qquad \qquad (4)
\]
For open boundary conditions, the mean density of ''go and stop''
vehicles per time step reads
\[
\rho _{gs}=n_0-n_0^2=\beta (1-\beta ),\qquad \qquad (5)
\]
where $n_0=N_0/N$ is the fraction of the stopped vehicles, and $N_0$
is the number of stopped vehicles on the road.

As a consequence, energy dissipation rate $E_d$ in the case of
$v_{\max }=1$ can be obtained as
\[
E_d=\frac m2(\beta -\beta ^2).\qquad \qquad (6)
\]
In formula (6), energy dissipation rate $E_d$ is directly related to
the probability for a vehicle moving out of the system or the
probability for a vehicle occupying the last site of the system. As
shown in figure 3, theoretical analysis is in good agreement with
numerical results.

For $v_{\max }=1$ and $\alpha <1,$ a discontinuous change in $E_d$
also occurs at a critical point $\beta _{cd}$, and the position of
$\beta _{cd}$ shifts towards a high value of $\beta $ with
increasing $\alpha ,$ which is similar to that in the case of
$v_{\max }=5$.

\subsection{Effects of the boundary conditions on energy dissipation in the
nondeterministic case}

Next, we investigate the rate of energy dissipation $E_d$ when the
stochastic braking behaviours of drivers are considered, i.e.,
$p\neq 0$. In the nondeterministic case, the dissipated energy has
two different contributions: one coming from the stochastic noise
and one from the interactions between vehicles (this in fact is the
only reason for dissipating energy in the deterministic case). Let
$E_{di}$ and $E_{dr}$ denote the rate of energy dissipation caused
by the interactions and randomizations, respectively. And energy
dissipation rate $E_d=E_{di}+E_{dr}. $ Figure 4  and 5 show the
relation of the energy dissipation rate $E_{di}$ and $E_{dr}$ to the
extinction rate $\beta $ with different values of the
injection rate $\alpha $, respectively, in the case of $v_{\max }=5$ and p$%
=0.5.$ As shown in figure 4, there is a critical value of the
extinction rate $\beta _{cr}$ below which $E_{di}$ increases with
increasing $\beta $, but above which $E_{di}$ abruptly decreases,
except for the case of $\alpha =1.$ Different from $E_{di}$, energy
dissipation rate $E_{dr}$ sharp increases at the critical point
$\beta _{cr}$ above which $E_{dr}$ shows the approximate plateau and
is independent of $\beta ,$ except for $\alpha =1,$ as shown in
figure 5. And the position of $\beta _{cr}$ shifts towards a high
value of $\beta $ with the increase of the injection rate $\alpha $.
For $\alpha =1,$ there is no discontinuous change in $E_{di}$ and
$E_{dr}$. Compared figure 4  with 5, it turns out that energy
dissipation is mainly caused by the randomization in the low-density
phase, and by the interactions in the high-density phase. When
$\alpha >0.35,$ the values of
the rate of energy dissipation $E_{di}$ and $E_{dr}$ for various value of $%
\alpha $ collapse into a single curve (not shown).

\begin{figure}
\includegraphics[height=5cm]{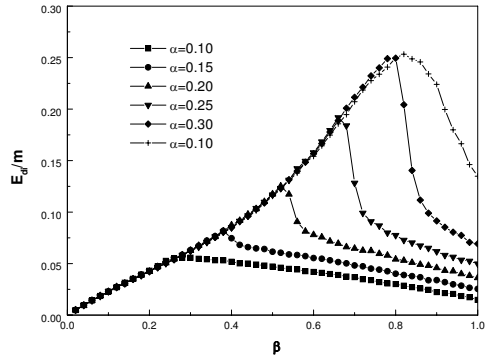}
\caption{\label{fig:epsart} Energy dissipation rate $E_{di}$ (scaled
by $m$ ) as a function of the extinction rate $\beta $ in the
non-deterministic NaSch model with $v_{\max }=5$ and $p=0.5$ for
various values of the injection rate $\alpha $.}
\end{figure}

\begin{figure}
\includegraphics[height=5cm]{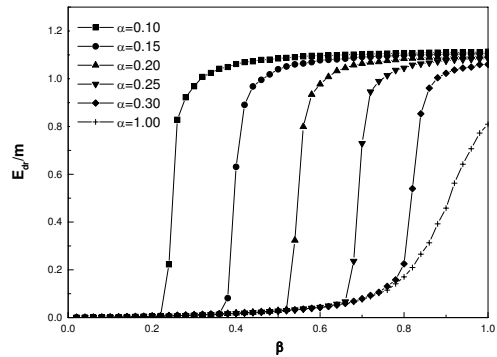}
\caption{\label{fig:epsart} Energy dissipation rate $E_{dr}$ (scaled
by $m$ ) as a function of the extinction rate $\beta $ in the
non-deterministic NaSch model with $v_{\max }=5$ and $p=0.5$ for
various values of the injection rate $\alpha $.}
\end{figure}

In fact, in the case of p$=0.5$ and $v_{\max }=5,$ there are the
transitions from the high-density to low-density phase for $\alpha
\leqslant 0.35$ and from high-density to maximum current phase for
$\alpha >0.35$\cite{29}. In the high-density phase, with the
increase of the extinction rate $\beta $, the mean velocity
increases and the dissipated energy increases. In the low-density
and maximum current phase, with increasing the rate $\beta $, the
distance headways increases and the interactions lowers; thus energy
dissipation rate $E_{di}$ decreases. Energy dissipation rate
$E_{dr}$ reaches a constant value in the low-density phase, but
continuatively increases with the increase of $\beta $ in the
maximum current phase.

It should be noted that energy dissipation rate $E_d$ for $\alpha
=1$ is minimum, i.e., energy dissipation in the maximum current
phase is lower than that in the low-density phase (not shown).
Traffic flow, however, in the maximum current phase is maximal.

The relationship of $E_d$ to the extinction rate $\beta $ for
different values of $v_{\max }$ in the case of $\alpha =1$ is shown
in figure 6. Similar to the deterministic case, when the value of
$\beta $ is very small, the rate of energy dissipation shows a
scaling relation and is independent
of the speed limit, as shown in figure 6. In the region of high values of $%
\beta $, the scaling relations of energy dissipation $E_d$ to the
speed limit cannot be observed; and $E_d$ increases with the
increase of $v_{\max } $. When $v_{\max }>3$, energy dissipation
rate $E_d$ increases with increasing the rate $\beta $ in the region
of high values of $\beta $. However, when $v_{\max }\leqslant \ 3$,
the value of $E_d$ tends to be invariable. Though the states of the
system for different speed limit are maximum current phase in the
region of high values of $\beta $, with the increase of $\beta $,
the mean velocity increases for $v_{\max }>3$ but does not vary in
the case of $v_{\max }\leqslant \ 3$. Consequently, there is a
plateau for the case $v_{\max }\leqslant \ 3$ in the region of high
values of $\beta $, which is different from that in the case of
$v_{\max }>3$.

\begin{figure}
\includegraphics[height=5cm]{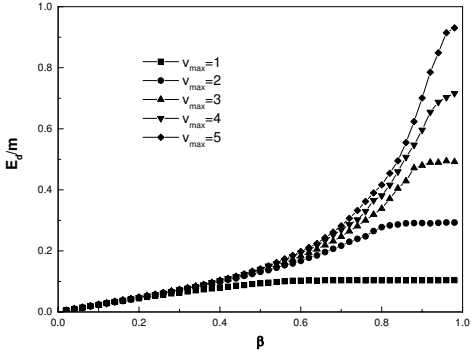}
\caption{\label{fig:epsart} Energy dissipation rate $E_d$ (scaled by
$m$ ) as a function of the extinction rate $\beta $ in the
non-deterministic NaSch model with $\alpha =1.0$ and $p=0.5$ for
various values of the speed limit $v_{\max }$.}
\end{figure}

In the system with $v_{\max }=1$, some vehicles can be stopped due
to the stochastic braking, therefore the rate of energy dissipation
$E_d$ is proportional to the mean ''go and stop'' density $\rho
_{gs},$ which demonstrates that the probability for ''go and stop''
vehicle to appear per time step in the system.

In the low-density phase A ($\alpha <\beta ,$ $\alpha <\alpha _c=1-\sqrt{p}%
), $ the fraction of the stopped vehicles reads

\[
n_0=1-\frac{q-\alpha }{1-\alpha },\qquad \qquad (7)
\]
where $q=1-p.$ And the mean ''go and stop'' density $\rho _{gs}$ can
be obtain as
\[
\rho _{gs}=n_0-n_0^2=\frac{q-\alpha }{1-\alpha }\left( 1-\frac{q-\alpha }{%
1-\alpha }\right) .\qquad \qquad (8)
\]
Substituting formula (8) into (4), we can obtain energy dissipation rate $%
E_d $ in the low-density phase A

\[
E_d^A=\frac{m(q-\alpha )(1-q)}{2(1-\alpha )^2}.\qquad \qquad (9)
\]

Figure 7 shows the relation between the rate of energy dissipation
$E_d$ and
the injection rate $\alpha $ with various values of $p$, in the case of $%
\beta =1.$ As shown in Fig. 7 in the region of low values of $\alpha
$, formula (9) gives good agreement with the simulation data.

\begin{figure}
\includegraphics[height=5cm]{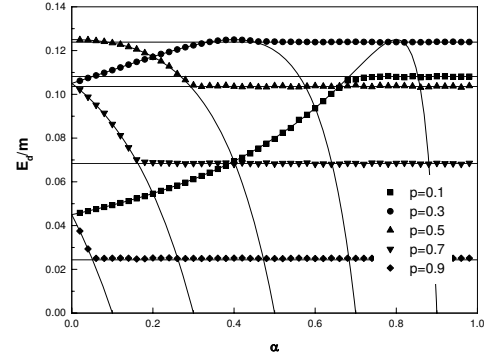}
\caption{\label{fig:epsart} Energy dissipation rate $E_d$ (scaled by
$m$ ) as a function of the injection rate $\alpha $ in the
non-deterministic NaSch model in the case of $v_{\max }=1$ and
$\beta =1.0$ for various values of the stochastic braking
probability $p$. Symbol data are obtained from computer simulations,
and solid line corresponds to analytic results of the formula (9)
and (15).}
\end{figure}

\begin{figure}
\includegraphics[height=5cm]{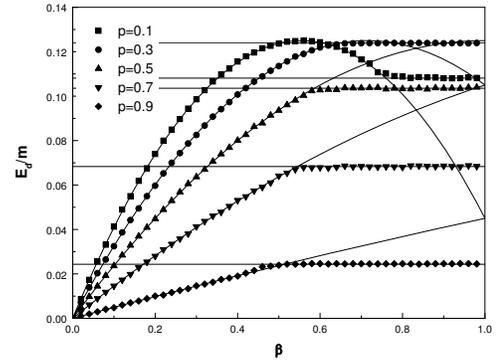}
\caption{\label{fig:epsart} Energy dissipation rate $E_d$ (scaled by
$m$ ) as a function of the extinction rate $\beta $ in the
non-deterministic NaSch model in the case of $v_{\max }=1$ and
$\alpha =1.0$ for various values of the stochastic braking
probability $p$. Symbol data are obtained from computer simulations,
and solid line corresponds to analytic results of the formula (12)
and (15).}
\end{figure}

From $\frac{dE_d^A}{d\alpha }=0,$ we obtain the critical value
$\alpha _{mc}=1-2p$ in which the curve described by formula (9)
reaches the maximum. Compared with the critical rate $\alpha
_c=1-\sqrt{p},$ the critical value of the stochastic braking
probability $p_c^L=\frac 14$ can be obtained. When $p<p_c^L,$ energy
dissipation rate $E_d^A$ increases with increasing the injection
rate $\alpha .$ When $p\geqslant 0.5,$ however, with the increase of
the rate $\alpha ,$ energy dissipation rate $E_d^A$ decreases. In
the interval $0.5>p\geqslant p_c^L,$ with increasing $\alpha ,$ the
rate of energy dissipation $E_d^A$ increases first and decreases
after a maximum value is reached.

In the high-density phase B ($\beta <\alpha ,$ $\beta <\beta _c=\frac 1{1+%
\sqrt{p}}$), the model of this paper is different from the ASEP with
parallel update and the mean velocity is determined not only by the
extinction rate $\beta ,$ but also the stochastic braking
probability $p.$ The fraction of the stopped vehicles reads

\[
n_0=1-q\beta .\qquad \qquad (10)
\]
And the mean ''go and stop'' density $\rho _{gs}$ can be written as
\[
\rho _{gs}=n_0-n_0^2=q\beta (1-q\beta ).\qquad \qquad (11)
\]
Substituting formula (11) into (4), we can obtain the rate of energy
dissipation $E_d$ in the high-density phase B

\[
E_d^B=\frac m2(q\beta -q^2\beta ^2).\qquad \qquad (12)
\]

Figure 8 shows the relation the energy dissipation rate $E_d$ as a
function of the extinction rate $\beta $ with different values of
the stochastic braking probability $p,$ in the case of $\alpha =1.$
As shown in Fig. 8, the
agreement can be obtained in the case of low value of the extinction rate $%
\beta .$

From $\frac{dE_d^B}{d\alpha }=0,$ we can obtain the critical value
$\beta _{mc}=\frac 1{2(1-p)}$ in which the curve corresponding to
formula (12)
reaches the maximum. Compared with the critical rate $\beta _c=\frac 1{1+%
\sqrt{p}},$ the critical value of the stochastic braking probability $p_c^h=%
\frac 14$ can be obtained. When $p\geqslant p_c^h,$ energy dissipation rate $%
E_d^B$ increases with the increase of the extinction rate $\beta .$ When $%
p<p_c^h,$ however, with increasing the rate $\beta ,$ energy
dissipation rate $E_d^B$ increases first and decreases after a
maximum value is reached.

In the maximum current phase C ($\alpha >\alpha _c,\beta >\beta
_c$), the fraction of the stopped vehicles reads

\[
n_0=\sqrt{p}.\qquad \qquad (13)
\]
And the mean ''go and stop'' density $\rho _{gs}$ can be obtain as

\[
\rho _{gs}=n_0-n_0^2=\sqrt{p}(1-\sqrt{p}).\qquad \qquad (14)
\]
Substituting formula (14) into (4), we can obtain energy dissipation rate $%
E_d$ in the maximum current phase C

\[
E_d^C=\frac m2(\sqrt{p}-p).\qquad \qquad (15)
\]
Equation (15) demonstrates that energy dissipation in the maximum
current phase is independent of the rate $\alpha $ and $\beta ,$ and
is only determined by the stochastic braking probability. Figure 7
and 8 give a comparison between number results and Eq. 15. As shown
in the right region of figure 7 and 8, formula (15) gives good
agreement with the simulation data.

\section{SUMMARY}

In this paper, we investigate the rate of energy dissipation caused
by braking in the NaSch model with open boundary conditions.
Different from periodic systems, open systems in which the vehicle
density is only a derived parameter are controlled by the injection
and extinction rate. In fact, real traffic systems are usually open,
hence it is highly desirable to investigate energy dissipation in
traffic systems both numerically and theoretically.

Numerical results show that in the deterministic case there is a
critical value of the extinction rate $\beta _{cd}$ above which
$E_d$ decreases
abruptly for $v_{\max }\geqslant 3,$ however, no discontiguous change in $%
E_d $ occurs when $v_{\max }<3$. The rate $\beta _{cd}$ is related
not only to the injection rate $\alpha ,$ but also to the maximum
velocity of vehicles. In the nondeterministic case, there is also a
critical value of the extinction rate $\beta _{cr}$ below which
$E_{di}$ and $E_{dr}$ increase
with increasing $\beta ,$ above which $E_{di}$ abruptly decreases but $%
E_{dr} $ sharp increases and shows the approximate plateau with
further increase of $\beta $, when the transition from the
high-density to low-density phase occurs. However, when the
transition from the high-density
to the maximum current phase occurs, the values of energy dissipation rate $%
E_{di}$ and $E_{dr}$ for various value of $\alpha $ collapse into a
single curve, and no discontinuous change occurs. Moreover$,$ there
is a nonvanishing energy dissipation rate, and no ''true'' free-flow
phase exists in the deterministic and nondeterministic NaSch models
with open boundary conditions.

Energy dissipation rate $E_d$ is directly related to traffic phase.
Energy dissipation in maximum current phase is smaller than that in
the low-density phase. A phenomenological mean-field theory is
presented to describe the energy dissipation rate $E_d$ in three
phases (low-density, high-density and maximum current phase) in the
case of $v_{\max }=1$. Theoretical analyses
give an excellent agreement with numerical results. But in the case of $%
v_{\max }>1$, explicit expressions about the energy dissipation rate
$E_d$ do not be obtained because of effects of long length of time
space correlations, and deserve further investigate.

\end{document}